\documentclass[12pt,preprint]{aastex}
\usepackage{epsfig}

\shorttitle{CLUSTER GALAXY ASSEMBLY TIMESCALE}
\shortauthors{CARRETERO ET AL.}

\begin{document}

\title{On the environmental dependence of cluster galaxy assembly timescale}

\author{C. Carretero\altaffilmark{1}, A. Vazdekis\altaffilmark{1},
J. E. Beckman\altaffilmark{1,}\altaffilmark{2}, P.
S\'anchez-Bl\'azquez\altaffilmark{3} and J.
Gorgas\altaffilmark{3}} 
\altaffiltext{1}{Instituto de
Astrof\'{\i}sica de Canarias, V\'{\i}a L\'actea s/n, 38200 La
Laguna, Tenerife, Spain; cch@iac.es} 
\altaffiltext{2}{Consejo
Superior de Investigaciones Cient\'{\i}ficas, Spain.}
\altaffiltext{3}{Departamento de Astrof\'{\i}sica, Facultad de
F\'{\i}sicas, Universidad Complutense de Madrid, 28040 Madrid, Spain.}

\begin{abstract}
We present estimates of CN and Mg overabundances with respect to
Fe for early-type galaxies in 8 clusters over a range of richness
and morphology. Spectra were taken from the Sloan Digital Sky
Survey (SDSS) DR1, and from WHT and CAHA
observations. Abundances were derived from absorption lines and
single burst population models, by comparing galaxy spectra with
appropriately broadened synthetic model spectra. We detect
correlations between [Mg/CN] and [CN/Fe] and
cluster X-ray luminosity. No correlation is observed for [Mg/Fe].
We also see a clear trend with the richness and morphology of the
clusters. This is interpreted given varying formation
timescales for CN, Mg and Fe, and a varying star
formation history in early-type galaxies as a function of their
environment: intermediate-mass early-type galaxies in more massive clusters are assembled on shorter timescales than in less massive clusters, with an upper limit of $\sim1$~Gyr.
\end{abstract}


\section{Introduction}

A key question for scenarios of galaxy formation is whether galaxies formed in
single structural ``monolithic'' events \citep{lar74} or by a series of
``hierarchical'' processes \citep{pre74,whi91} in which large galaxies built up from smaller ones. Many structural and dynamical properties of
galaxies in clusters are explained in this scenario, though problems remain: the
absence of the predicted mass cusps in the centres of ellipticals and bulges, and the prediction of far more satellite galaxies than those observed.

Stellar populations offer a fossil record of the formation and evolution of
galaxies, most clearly in elliptical galaxies, and stellar 
population studies provide very strong
constraints on the principal galaxy formation scenarios. It is 
hard to reconcile the hierarchical models with the result that massive galaxies show significantly larger mean luminosity weighted ages than their smaller counterparts \citep{kau03}.

Understanding stellar populations in early-type
galaxies as a function of the environment can provide answers
to the puzzle. The present observational base is small. Only
three clusters have been observed for a detailed stellar
populations analysis: Virgo \citep[e.g.][]{vaz01a}, Coma
\citep[e.g.][]{jor99} and Fornax \citep{kun98}. Extending the
study to a large number of clusters covering a range of richness
and morphology is mandatory.

Past studies of clusters used the original Lick/IDS spectral
indices \citep{wor94b} whose reliability is limited by their
resolution dependence, as uncertain corrections for broadening and
instrumental effects are needed. It is better to use modelled
integrated spectra, which can be broadened to match galaxy
velocity dispersion, $\sigma$. Such models have been developed by
\citet{vaz99} and allow an accurate separation of age and
metallicity, and subsequent individual abundance derivations.

The study of the element abundance ratios in elliptical galaxies
within distinct clusters should be a powerful discriminant between
different star formation histories \citep[e.g.][]{wor98}. In
particular, overabundances of [Mg/Fe] compared with the solar
ratio have been found in massive elliptical galaxies
\citep{pel89,wor92,vaz97}. These have been interpreted via several possible scenarios based on the fact that Mg is
mainly produced in Type II supernovae \citep{fab92,mat94}, and include different star formation rates (SFR) and a time dependent IMF.

Differences in the abundances of C and N as a function of the
environment have been recently suggested by \citet{pat03}, who found striking spectral differences between field elliptical
galaxies and their counterparts in the central region of the Coma
cluster. Galaxies in the denser environment showed
significantly lower CN$_2$ and C4668 absorption strength. Here we 
explore these differences by extending the
study to a larger number of clusters over a range of richness and
morphology, applying the new analysis techniques to derive
abundance ratios. We have assumed a flat Universe with H$_0 =
75$~km~s$^{-1}$~Mpc$^{-1}$ and $q_0 = 0.5$.

\section{Data}

Sloan Digital Sky Survey \citep{sto02} spectra were obtained using
a multiobject, $3^{\prime\prime}$ diameter fiber spectrograph. Exposures ranged from 45 to 75 minutes. All the data processing
was performed with automated SDSS software. Redshifts were
measured on the reduced spectra by an automated system, which
models each galaxy spectrum as a linear combination of stellar
populations. We measured independently the redshifts of the
galaxies used in this study by crosscorrelating each galaxy
spectrum with our SSP synthetic model spectra. We found no
significant differences between SDSS redshift values and ours.

From the SDSS Data Release 1 database
we selected galaxy spectra according to the following criteria. They must:

{\it i) Belong to an Abell cluster.} We included this criterion
because the richness and the morphology of Abell clusters are
uniformly defined and described in the literature. Also, X-ray
luminosity values are available.

{\it ii) Belong to the early-type galaxies catalogue of
\citet{ber03}.} This sample has $\sim9000$ early-type galaxies,
in the redshift range $0.01\le z\le 0.3$, selected from the SDSS
spectroscopic database using morphological and spectral criteria.
The mean spectrum signal-to-noise per pixel is 16.

{\it iii) Have $S/N$ per pixel greater than 15.}

{\it iv) Have velocity dispersion in the range 150~km~s$^{-1}$
$\le\sigma\le$ 250~km~s$^{-1}$.} Galaxies outside this range of
$\sigma$ were rejected because of the completeness of the sample:
not all clusters had spectra of dwarf and/or giant elliptical
galaxies because of the inner limitation of SDSS data (for dwarfs)
and the morphology of the clusters (for giants). Also, the quality of the spectra of the faintest galaxies was too low for our analysis requirements.
$\sigma$ values  were obtained from \citet{ber03}.

Using these criteria, we obtained a total of 55
galaxies distributed in 6 clusters. The clusters are: A257, A279,
A655, A1238, A1650 and A2050. Their redshift values vary in the
range $0.07 < z < 0.13$ and they cover a range of richness and morphology. 
See details in Table~\ref{table1}.

For comparison and completeness, we added to our SDSS data
high-quality long-slit spectra of early-type galaxies in Coma and
Virgo clusters \citep[for details, see][]{pat03}. To compare them with those of the SDSS sample, we extracted spectra along the
slit simulating a circular aperture (distance weighted coadded
spectra) of radius $1.5^{\prime\prime}$ at $z = 0.1$.
This aperture translates into apertures of radius 
$6^{\prime\prime}$ for Coma and $37^{\prime\prime}$ for Virgo.

\section{Galaxy measurements and results}

To derive mean luminosity-weighted ages and metallicities, we 
compared selected absorption line strengths with those predicted
by the model of \citet{vaz99}. This model provides flux-calibrated
spectra in the optical range at a resolution of 1.8~\AA\ (FWHM)
for single-burst stellar populations. This way, we can transform synthetic spectra to the resolution and
dispersion of the galaxy spectra instead of the opposite, as
required while working in the Lick system. Selected absorption
indices were CN$_2$, Mg$_2$ \citep{wor94a} and Fe2 (defined as Fe2 = $\frac{{\rm Fe}4383+{\rm Fe}5270}{2}$). We used
these features because of their low sensitivity to variations in
$S/N$ \citep{car03} and velocity dispersion (we have estimated
 $\Delta(index)/index < 0.15$, for $\Delta(\sigma)$ =
300~km~s$^{-1}$). This way we avoid possible variations in the
index value, as $\sigma$ may vary as a function of $r$, due to the
fact that SDSS spectra provide the light integrated within the
fibers of the spectrograph.

\begin{figure}
\figurenum{1}
\epsscale{0.8}
\plotone{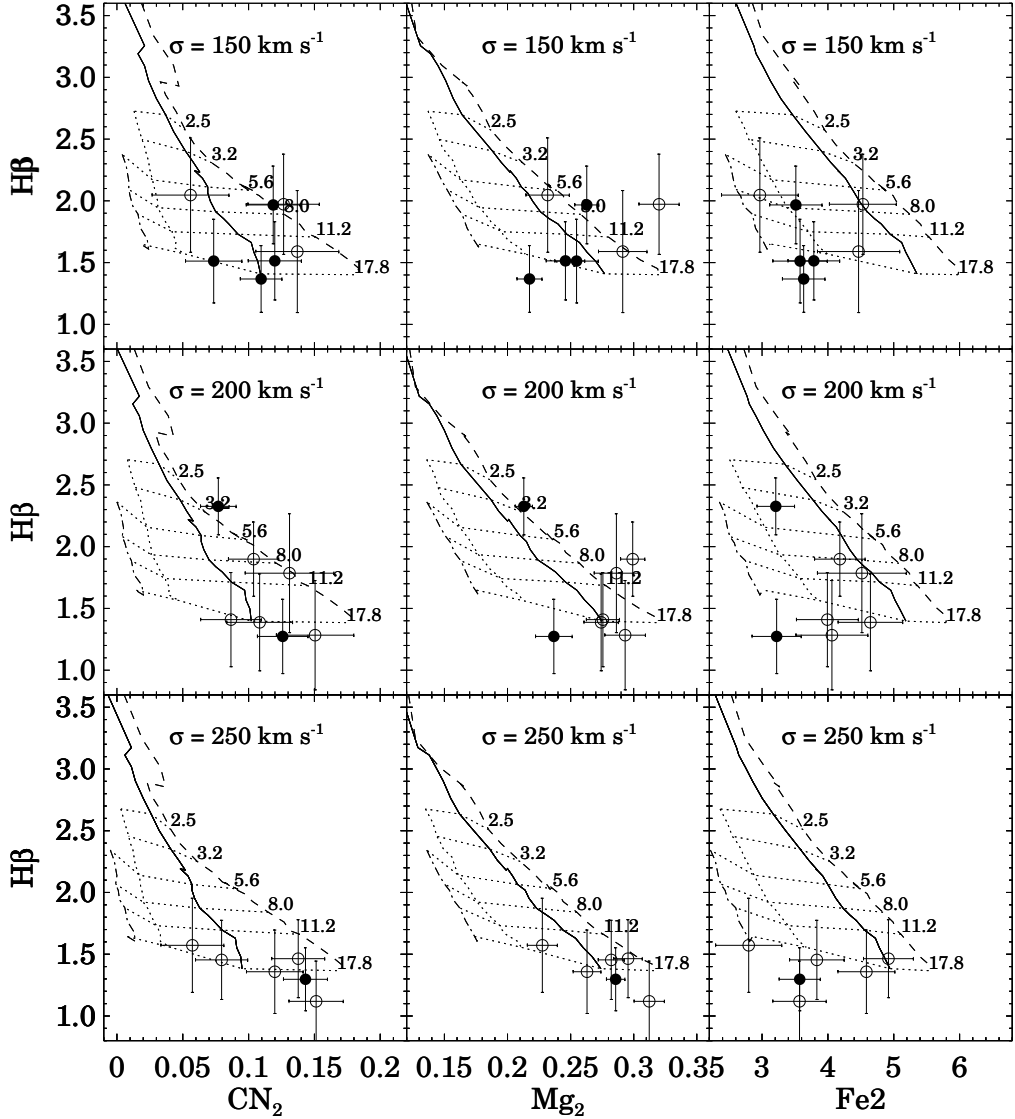}
\caption{Model grids
for the galaxies of clusters A1238 (filled circles) and A655 (open
circles). These two clusters have extreme values of X-ray
luminosity. {\it From left to right:} Plots of H$\beta$ vs. the
metallicity indices CN$_2$, Mg$_2$ and Fe2 (defined in the text).
The galaxy velocity dispersions increase from top to bottom, in
the three bins of $\sigma$ quoted in the boxes, to which galaxy
and model spectra have been broadened. Overplotted are the models
by \citet{vaz99}. Lines of constant [Fe/H] $=-0.7,-0.4,0.0$ and
$0.2$ are shown by dot-dashed, dotted, solid and dashed lines,
respectively. Thin dotted horizontal lines represent models of
constant age, quoted in gigayears. The ages become increasingly uncertain for
values greater than 8 Gyr \citep[see][]{wor94b,vaz99} but these uncertainties have no
significant effects on relative abundance determination.}
\label{fig1}
\end{figure}

Plots of the strengths of the selected indices versus H$\beta$
provide close to orthogonal model grids, allowing us to estimate accurately
 galaxy mean ages as well as relative abundances of the
different elements. Figure~\ref{fig1} illustrates this method for
the galaxies of clusters A1238 and A655 (two clusters with extreme
values of X-ray luminosity). We will refer to the metallicities
derived in the diagrams CN$_2$--H$\beta$, Mg$_2$--H$\beta$ and
Fe2--H$\beta$ as $Z_{\rm CN}$, $Z_{\rm Mg}$ and $Z_{\rm Fe}$,
respectively. Since the CN$_2$ index is strongly dominated by C
and N, the Mg$_2$ index is governed by Mg and the Fe2 index by Fe
\citep{tri95}, these metallicities must be close to the [CN/H],
[Mg/H] and [Fe/H] abundances, and $[Z_{\rm CN}/Z_{\rm Fe}]$,
$[Z_{\rm Mg}/Z_{\rm Fe}]$ and $[Z_{\rm Mg}/Z_{\rm CN}]$ are then
estimates of the abundance ratios [CN/Fe], [Mg/Fe] and [Mg/CN] for
each galaxy. Note that an extrapolation of the model grids is
required for some galaxies to obtain the abundances of CN and Mg,
since the models extend only to [M/H] = 0.2. It is worth noting
that certain galaxies fall below the model grids, which can be
attributed to the fact that absolute age determination is subject
to model uncertainties \citep[see][]{vaz01b,sch02}. We neglect the
possible effect of nebular emission on H$_\beta$ since no
[\ion{O}{3}]$\lambda$5007 emission is detected in our galaxy
spectra. Nevertheless, assuming an upper H$_\beta$ emission
correction of $\sim0.5$~\AA\ \citep{kun02}, although this
would give rise to a significant reduction in the mean age of the
stellar populations of the oldest galaxies, the net effect on the
abundance ratios would be no more than $\sim0.05$ dex, for the
most affected galaxies in a cluster. Note that this correction for the 
H$\beta$ index is larger than the one obtained by
varying the model prescriptions, e.g. with $\alpha$-enhanced isochrones and
atomic diffusion included \citep{vaz01b}, for inferring ages in
better agreement with the current age of the Universe, for the oldest galaxies
in our sample. However we do recognize that the ages quoted here, and in current articles dealing with stellar population modelling, may well span an older range than the true age range of the populations observed. This would indicate some flaw or flaws in the present quantitaive understanding of stellar evolution, and is a problem recognized by the community working in population synthesis.

\begin{figure}
\figurenum{2} 
\epsscale{0.8} 
\plotone{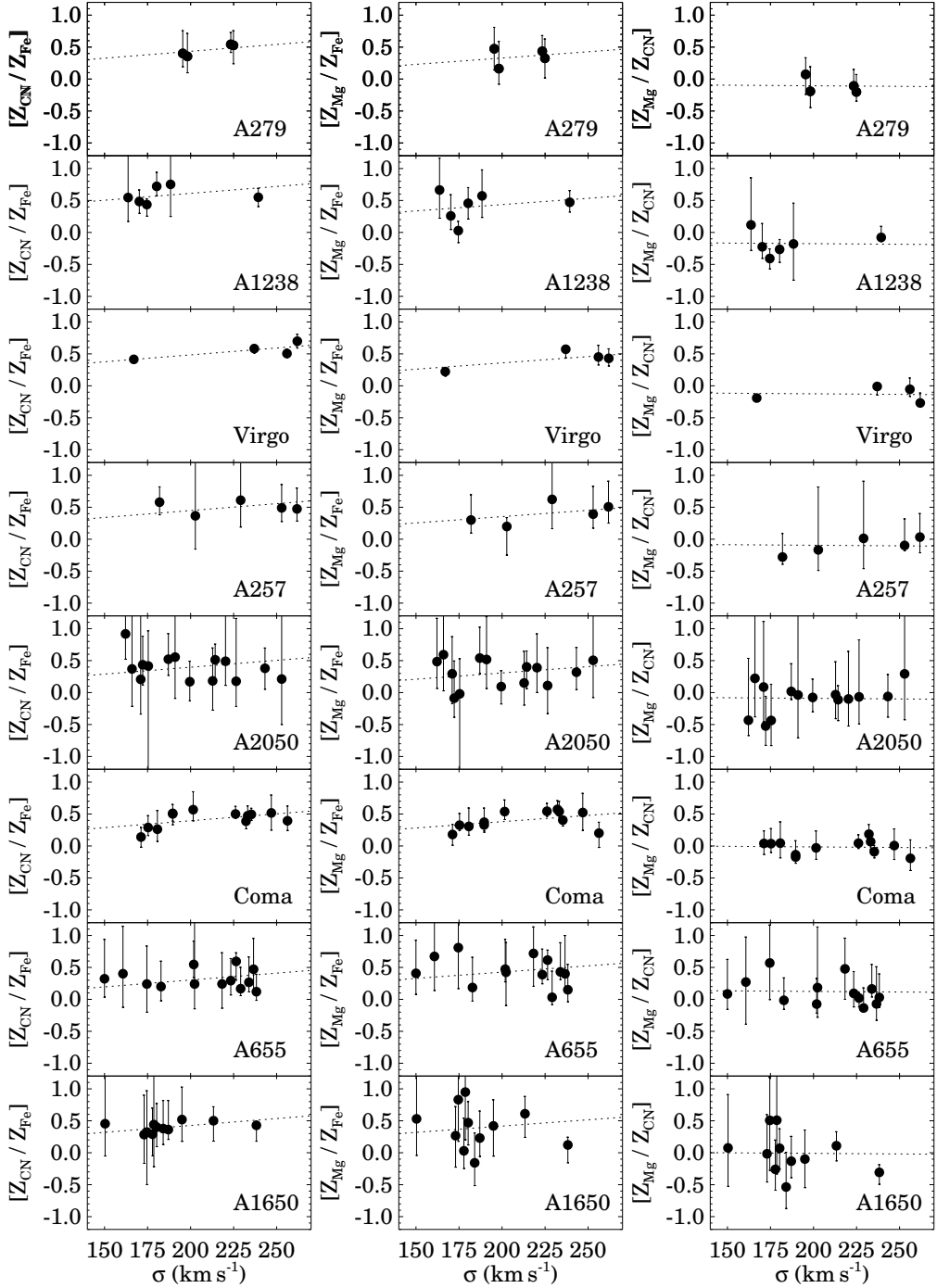}
\caption{Abundance
ratios $[Z_{\rm CN}/Z_{\rm Fe}]$ {\it (left)}, $[Z_{\rm Mg}/Z_{\rm
Fe}]$ {\it (centre)} and $[Z_{\rm Mg}/Z_{\rm CN}]$ {\it (right)}
versus galaxy velocity dispersion, $\sigma$. Each point
corresponds to one galaxy within a cluster.  Errors are
computed from the measured index error bars shown in
Fig.~\ref{fig1}. Galaxy clusters are ordered by decreasing X-ray
luminosity, from top to bottom. The dotted straight lines show the
level of relative abundances for each cluster. See the text for
more details.}
\label{fig2}
\end{figure}

Figure~\ref{fig2} shows the measured abundance ratios for the
galaxies in the whole sample of clusters as a function of
the velocity dispersion. To derive representative
relative abundances ratios for each cluster, and since relative
abundances correlate with $\sigma$, we have fitted to each cluster
a straight line with a fixed slope and a varying intercept. We
have assumed that this slope corresponds to that found for the
Coma cluster in the considered range of $\sigma$. Note that we do
not have any, statistically significant, evidence of a variation
of the slope between the different clusters. In Table~\ref{table1}
we list the derived relative abundances at a fixed velocity
dispersion of $\sigma=200$~km~s$^{-1}$ for each cluster.

As a quantitative indicator of the masses of the clusters, we have
used X-ray luminosities. $L_{\rm X}$ values were taken from
\citet{ebe98} and \citet{led03}, adapted to the cosmological
parameters used by the latter.

Figure~\ref{fig3} shows the values of $[Z_{\rm CN}/Z_{\rm Fe}]$,
$[Z_{\rm Mg}/Z_{\rm Fe}]$ and $[Z_{\rm Mg}/Z_{\rm CN}]$ versus
X-ray luminosity for each cluster. We find clear correlations
between [CN/Fe] and [Mg/CN] values and X-ray luminosity, with probabilities of
no correlation of 0.046 and 0.002, respectively. On the
contrary, no correlation is found for [Mg/Fe]. It is worth noting
that the significant point in these relations is the relative
differences in abundance ratios, not their absolute values. It is
noticeable that the abundance ratio values also correlate with the
richness class and the morphological type (see Table~\ref{table1}), with 
probabilities of no correlation of 0.05 and 0.02, respectively.

\section{Discussion}

The correlations can be interpreted in terms of the
different formation timescales for each element, and the different
star formation histories of early-type galaxies as a function of
their environment.

Magnesium is ejected into the interstellar medium (ISM) by
Type II supernovae (SNe II) on short timescales ($<10$~Myr). On
the other hand, the iron-peak elements are the products of SNe~Ia,
which occur on timescales of $\sim1$~Gyr. Between the two
extremes, although there are recent suggestions that most of the C
come from massive stars \citep{ake04}, C and N are mainly ejected
into the ISM by low- and intermediate-mass stars
\citep{ren81,chi03}, leading to CN formation on timescales longer
than for Mg but shorter than for Fe. Furthermore, several authors
\citep[e.g.][]{ell97,sta98} argue that early-type galaxies are old
and passively evolving systems. In any case, the
luminosity-weighted ages derived from our model grids confirm that
the galaxies are significantly older than the formation timescales
of the different species. So, if we find substantial differences
in the abundance ratios of these elements which depend on the
physical properties of the environment, these must be due to the
fact that galaxies are assembled on different timescales as a
function of their environment.

\begin{figure}
\figurenum{3} 
\epsscale{0.8} 
\plotone{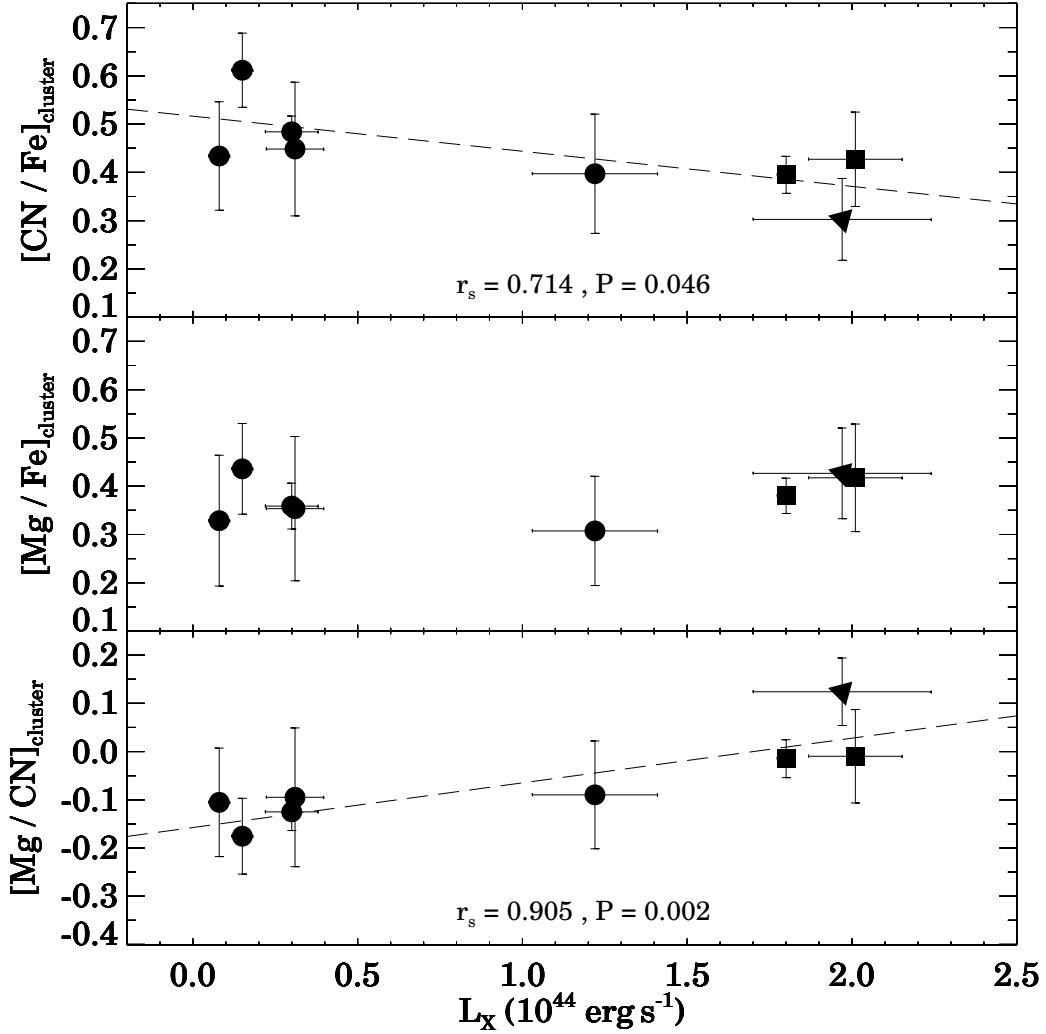} 
\caption{Cluster
X-ray luminosity versus overabundance values of $[Z_{\rm
CN}/Z_{\rm Fe}]$ {\it(top)}, $[Z_{\rm Mg}/Z_{\rm Fe}]$
{\it(middle)} and $[Z_{\rm Mg}/Z_{\rm CN}]$ {\it(bottom)}.
Circles, squares and triangles indicate clusters with richness
classes of 1, 2 and 3, respectively. Each point corresponds to one
individual cluster and represents the interpolated abundance ratio
for $\sigma=200$~km~s$^{-1}$. The Spearman rank-order correlation
coefficients and its significance values are written in top and
bottom panels.} 
\label{fig3}
\end{figure}

In this framework, the constancy of the [Mg/Fe] 
values is explained in terms of the great difference in the
formation timescales of the two elements: the galaxies are fully
assembled before Type Ia SNe can significantly pollute with Fe the
ISM of the smaller galaxies before merging, and right after Mg is
fully ejected. Since [Mg/Fe] is found to be constant with the
X-ray luminosity of the clusters (see Fig.~\ref{fig3}), which is
an indicator of their mass, we conclude that this ratio is
independent of the environment. Similar results for the [Mg/Fe]
ratio have been obtained by other authors
\citep{jor99,kun02,pat03}, by studying field and Coma cluster
elliptical galaxies.

However, when considering species with less disparate formation
timescales, such as CN and Fe, or CN and Mg, clear correlations
are found between abundance ratios and the environment, as shown
in Fig.~\ref{fig3}. The fact that [CN/Fe] decreases with the
cluster X-ray luminosity, and that [Mg/CN] increases with it,
suggests that galaxies in more massive clusters are fully
assembled on shorter timescales than those in less massive
clusters. We show that this difference is large enough to produce measurable
variations of the abundance ratios of galaxies in more or less
massive clusters.

The result that there exist relative differences in the assembly timescales of
the galaxies due to the properties of the environment is qualitatively in agreement with the
hierarchical models. Discrepancies appear, however, when considering the absolute values
of such timescales. The fact that [CN/Fe] abundance ratio is not constant implies
that early-type galaxies are fully assembled on timescales around the massive
release of CN into the ISM. Hierarchical models, on the
contrary, predict longer assembly timescales.

Other scenarios have been explored in order to explain the
differences in abundance ratio values as functions of the
environment \citep[see][]{pat03}. These include
a decrease in the stellar giant/dwarf ratio in high-density
environments, with respect to low-density ones, which would lead
to lower index values in the latter. But model calculations have
shown that the differences due to a variation in the IMF are too small to produce the observed variations. Also, a
difference in the luminosity-weighted mean age between high- and
low-density environments has been proposed. Models
show that, to account for the differences, the galaxies
in high-density environments must be $\sim8$~Gyr younger than in
 low-density ones. This contradicts previous studies that
suggest that galaxies in high-density environments are, in any
case, older than those in low-density environments \citep{kun02}. So our interpretation here seems the most consistent.

It is noteworthy that we have found a dependence of the
abundance ratios with several properties of the environment, both
quantitative (i.e. X-ray luminosity values) and qualitative (i.e.
richness and morphological types) as shown in Table~\ref{table1}.
This lends strong support to the basic hypothesis that the
characteristics of the environment affect the evolution of the
galaxies. The relations we have found set clear constraints on
models of chemical evolution and galaxy formation.

\acknowledgements
The authors thank A. Aguerri, X. Barcons, L. Carigi, C. Guti\'errez, R. Peletier and the anonymous referee for useful comments. We acknowledge grant AYA2001-0435 from the Spanish Ministry of Science and Technology. The SDSS is managed by the ARC for the Participating Institutions.

\begin{deluxetable}{ccccccc}
\tablecolumns{7} 
\tablecaption{Properties and measurements of
analyzed clusters} 
\tablehead{ \colhead{Cluster} & \colhead{$z$} &
\colhead{Rich.\tablenotemark{a}} &
\colhead{Morph. $^{\rm b}$} & \colhead{$L_{\rm X}$ $^{\rm c}$
} & \colhead{$[Z_{\rm CN}/Z_{\rm Fe}]$} &
\colhead{$[Z_{\rm Mg}/Z_{\rm Fe}]$}} 
\startdata A0279 & 0.080 & 1
& II--III & 0.08 & 0.43 & 0.33\\ A1238 & 0.073 & 1 & III     &
0.15 &  0.61 & 0.44\\ Virgo & 0.004 & 1 & III     & 0.30 &  0.48 &
0.36\\ A0257 & 0.070 & 1 & II--III & 0.31 &  0.45 & 0.35\\ A2050 &
0.118 & 1 & II--III & 1.22 &  0.40 & 0.31\\ Coma  & 0.023 & 2 & II
& 1.80 &  0.39 & 0.38\\ A0655 & 0.127 & 3 & I--II   & 1.97 &  0.30
& 0.43\\ A1650 & 0.085 & 2 & I--II   & 2.01 &  0.40 &~~0.39
\enddata
\tablenotetext{a}{Abell richness class. $^{\rm b}$~Bautz-Morgan morphological type. $^{\rm c}$~X-ray luminosity in units of 10$^{44}$ erg~s$^{-1}$.}
\label{table1}
\end{deluxetable}

\begin{thebibliography}{}

\bibitem[Akerman et al.(2004)]{ake04}Akerman, C.J., Carigi, L., Nissen, P.E.,
Pettini, M., \& Asplund, M. 2004, \aap, 414, 931
\bibitem[Bernardi et al.(2003)]{ber03}Bernardi, M. et al. 2003, \aj, 125, 1817
\bibitem[Cardiel et al.(2003)]{car03}Cardiel, N., Gorgas, J., S\'anchez-Bl\'azquez,
P., Cenarro, A.J., Pedraz, S., Bruzual, G., \& Klement, J. 2003, \aap, 409, 511
\bibitem[Chiappini, Romano \& Matteucci(2003)]{chi03}Chiappini, C., Romano, D., \&
Matteucci, F. 2003, \mnras, 339, 63
\bibitem[Ebeling et al.(1998)]{ebe98}Ebeling, H., Edge, A.C., B\"{o}hringer, H., Allen,
S.W., Crawford, C.S., Fabian, A.C., Voges, W., \& Huchra, J.P. 1998, \mnras, 301,
881
\bibitem[Ellis et al.(1997)]{ell97}Ellis, R.S., Smail, I., Dressler, A., Couch,
W.J., Oemler, A.J., Butcher, H., \& Sharples, R.M. 1997, \apj, 483, 582
\bibitem[Faber, Worthey, \& Gonz\'alez(1992)]{fab92}Faber, S.M., Worthey, G., \& Gonz\'alez, J.J.
1992, in IAU Symp. 149, Stellar Populations of Galaxies, ed. B. Barbury \& A.
Renzini (Dordrecht: Kluwer), 255
\bibitem[J{\o}rgensen(1999)]{jor99}J{\o}rgensen, I. 1999, \mnras, 306, 607
\bibitem[Kauffmann et al.(2003)]{kau03}Kauffmann, G. et al. 2003, \mnras, 341, 54
\bibitem[Kuntschner \& Davies(1998)]{kun98}Kuntschner, H., \& Davies, R. \mnras,
295, 29
\bibitem[Kuntschner et al.(2002)]{kun02}Kuntschner, H., Smith, R., Colles, M.,
Davies, R., Kaldare, R., \& Vazdekis, A. 2002, \mnras, 337, 172
\bibitem[Larson(1974)]{lar74}Larson, R.B. 1974, \mnras, 166, 585
\bibitem[Ledlow et al.(2003)]{led03}Ledlow, M.J., Voges, W., Owen, F.N., \& Burns,
J.O. 2003, \aj, 126, 2740
\bibitem[Matteucci(1994)]{mat94}Matteucci, F., 1994, \aap, 288, 57
\bibitem[Peletier(1989)]{pel89}Peletier, R.F. 1989, Ph.D. thesis, Univ. Groningen
\bibitem[Press \& Schechter(1974)]{pre74}Press, W.H., \& Schechter, P. 1974,
\apj, 187, 425
\bibitem[Renzini \& Voli(1981)]{ren81}Renzini, A., \& Voli, M. 1981, \aap, 94, 175
\bibitem[S\'anchez-Bl\'azquez et al.(2003)]{pat03}S\'anchez-Bl\'azquez, P., Gorgas,
J., Cardiel, N., Cenarro, J., \& Gonz\'alez, J.J. 2003, \apjl,
590, L91
\bibitem[Schiavon et al.(2002)]{sch02}Schiavon, R.P., Faber, S.M., Castilho,
B.V., \& Rose, J.A. 2002, \apj, 580, 850
\bibitem[Stanford, Eisenhardt, \& Dickinson(1998)]{sta98}Stanford, S.A.,
Eisenhardt, P., \& Dickinson, M. 1998, \apj, 492, 461
\bibitem[Stoughton et al.(2002)]{sto02}Stoughton, C. et al. 2002, \aj, 123, 485
\bibitem[Tripicco \& Bell(1995)]{tri95}Tripicco, M., \& Bell, R.A. 1995, \aj,
110, 3035
\bibitem[Vazdekis(1999)]{vaz99}Vazdekis, A. 1999, \apj, 513, 224
\bibitem[Vazdekis et al.(2001b)]{vaz01b}Vazdekis, A., Kuntschner, H., Davies, R.,
Arimoto, N., Nakamura, O., \& Peletier, R. 2001b, \apjl, 551, L127
\bibitem[Vazdekis et al.(1997)]{vaz97}Vazdekis, A., Peletier,R.F., Beckman,J.E.,
\& Casuso,E. 1997, \apjs, 111, 203
\bibitem[Vazdekis et al.(2001a)]{vaz01a}Vazdekis, A., Salaris, M., Arimoto, N.,
\& Rose, J.A. 2001a, \apj, 549, 274
\bibitem[White \& Frenk(1991)]{whi91}White, S.D.M., \& Frenk, C.S. 1991, \apj,
379, 521
\bibitem[Worthey(1994)]{wor94b}Worthey, G. 1994, \apjs, 95, 107
\bibitem[Worthey(1998)]{wor98}Worthey, G. 1998, \pasp, 110, 888
\bibitem[Worthey et al.(1992)]{wor92}Worthey, G., Faber, S.M., \& Gonz\'alez, J.J.
1992, \apj, 398, 69
\bibitem[Worthey et al.(1994)]{wor94a}Worthey, G., Faber, S.M., Gonz\'alez, J.J., \&
Burstein, D. 1994 \apjs, 94, 687

\end{thebibliography}
\end{document}